\documentstyle[epsfig]{mn}

\newif\ifAMStwofonts
\def\gtorder{\mathrel{\raise.3ex\hbox{$>$}\mkern-14mu
             \lower0.6ex\hbox{$\sim$}}}
\def\ltorder{\mathrel{\raise.3ex\hbox{$<$}\mkern-14mu
             \lower0.6ex\hbox{$\sim$}}}

\ifoldfss
  \ifCUPmtlplainloaded \else
    \NewTextAlphabet{textbfit} {cmbxti10} {}
    \NewTextAlphabet{textbfss} {cmssbx10} {}
    \NewMathAlphabet{mathbfit} {cmbxti10} {} % for math mode
    \NewMathAlphabet{mathbfss} {cmssbx10} {} %  "   "    "
  \fi
  \ifAMStwofonts
    \ifCUPmtlplainloaded \else
      \NewSymbolFont{upmath} {eurm10}
      \NewSymbolFont{AMSa} {msam10}
      \NewMathSymbol{\upi}     {0}{upmath}{19}
      \NewMathSymbol{\umu}     {0}{upmath}{16}
      \NewMathSymbol{\upartial}{0}{upmath}{40}
      \NewMathSymbol{\leqslant}{3}{AMSa}{36}
      \NewMathSymbol{\geqslant}{3}{AMSa}{3E}

       \let\le=\leqslant
       \let\ge=\geqslant
    \fi
  \fi
\fi % End of OFSS

\ifnfssone
  \newmathalphabet{\mathit}
  \addtoversion{normal}{\mathit}{cmr}{m}{it}
  \addtoversion{bold}{\mathit}{cmr}{bx}{it}
  \newmathalphabet{\mathbfit} % math mode version of \textbfit{..}
  \addtoversion{normal}{\mathbfit}{cmr}{bx}{it}
  \addtoversion{bold}{\mathbfit}{cmr}{bx}{it}
  \newmathalphabet{\mathbfss} % math mode version of \textbfss{..}
  \addtoversion{normal}{\mathbfss}{cmss}{bx}{n}
  \addtoversion{bold}{\mathbfss}{cmss}{bx}{n}
  \ifAMStwofonts
    \ifCUPmtlplainloaded \else
      \UseAMStwoboldmath
      \makeatletter
      \new@mathgroup\upmath@group
      \define@mathgroup\mv@normal\upmath@group{eur}{m}{n}
      \define@mathgroup\mv@bold\upmath@group{eur}{b}{n}
      \edef\UPM{\hexnumber\upmath@group}
      \new@mathgroup\amsa@group
      \define@mathgroup\mv@normal\amsa@group{msa}{m}{n}
      \define@mathgroup\mv@bold\amsa@group{msa}{m}{n}
      \edef\AMSa{\hexnumber\amsa@group}
      \makeatother
      \mathchardef\upi="0\UPM19
      \mathchardef\umu="0\UPM16
      \mathchardef\upartial="0\UPM40
      \mathchardef\leqslant="3\AMSa36
      \mathchardef\geqslant="3\AMSa3E

       \let\le=\leqslant
       \let\ge=\geqslant
    \fi
  \fi
\fi % End of NFSS release 1

\ifnfsstwo
  \DeclareMathAlphabet{\mathbfit}{OT1}{cmr}{bx}{it}
  \SetMathAlphabet\mathbfit{bold}{OT1}{cmr}{bx}{it}
  \DeclareMathAlphabet{\mathbfss}{OT1}{cmss}{bx}{n}
  \SetMathAlphabet\mathbfss{bold}{OT1}{cmss}{bx}{n}
  \ifAMStwofonts
    \ifCUPmtlplainloaded \else
      \DeclareSymbolFont{UPM}{U}{eur}{m}{n}
      \SetSymbolFont{UPM}{bold}{U}{eur}{b}{n}
      \DeclareSymbolFont{AMSa}{U}{msa}{m}{n}
      \DeclareMathSymbol{\upi}{0}{UPM}{"19}
      \DeclareMathSymbol{\umu}{0}{UPM}{"16}
      \DeclareMathSymbol{\upartial}{0}{UPM}{"40}
      \DeclareMathSymbol{\leqslant}{3}{AMSa}{"36}
      \DeclareMathSymbol{\geqslant}{3}{AMSa}{"3E}

       \let\le=\leqslant
       \let\ge=\geqslant
    \fi
  \fi
\fi % End of NFSS release 2

\ifCUPmtlplainloaded \else
  \ifAMStwofonts \else % If no AMS fonts
    \def\upi{\pi}
    \def\umu{\mu}
    \def\upartial{\partial}
  \fi
\fi

\title[The Redshift Distribution of Type-Ia Supernovae: 
Progenitors and Cosmic Star Formation History] {The Redshift Distribution of Type-Ia Supernovae: 
Constraints on Progenitors and Cosmic Star Formation History}

\date{Accepted - .
      Received - ;}

\author[A. Gal-Yam and D. Maoz]{Avishay Gal-Yam$^{1,2}$ 
and Dan Maoz$^{1}$\\
$^{1}$ School of Physics \& Astronomy and Wise Observatory, 
Tel Aviv University, Tel Aviv 69978, Israel; 
avishay@wise.tau.ac.il\\
$^{2}$ Colton Fellow. \\}

\begin{document}

\maketitle

%\label{firstpage}

\begin{abstract}

We use the redshift distribution of type-Ia supernovae (SNe) discovered by the 
Supernova Cosmology Project to constrain the star formation history (SFH)
of the Universe and SN~Ia progenitor models. Given some of the 
recent determinations of the SFH, the observed
SN~Ia redshift distribution indicates a
long ($\gtorder 1 h^{-1}$ Gyr) mean delay time between the formation of a
stellar population and the explosion of some of its members as SNe~Ia. 
For example, if the Madau et al. (1998) SFH is assumed, the delay time $\tau$ 
is constrained to be $\tau \ge 1.7 (\tau \ge 0.7) h^{-1}$ Gyr at the $95\%(99\%)$
confidence level (CL). SFHs that rise at high redshift, 
similar to those advocated by Lanzetta 
et al. (2002), are inconsistent with the data at the $95\%$ CL 
unless $\tau > 2.5 h^{-1}$ Gyr.
Long time delays disfavor progenitor
models such as edge-lit detonation
of a white dwarf accreting from a giant donor, and the carbon core ignition of 
a white dwarf passing the Chandrasekhar mass due to accretion from a subgiant.
The SN~Ia delay may be shorter, thereby relaxing some of these
constraints, if the field star formation rate falls, between $z=1$ and the
present, less sharply than implied, e.g., by the original Madau plot. We show
that the discovery of larger samples of high-$z$ SNe~Ia by forthcoming
observational projects should yield strong constraints on the progenitor 
models and the SFH.   
In a companion paper, we demonstrate that if SNe~Ia produce most 
of the iron in galaxy clusters, and the stars in clusters formed 
at $z\sim2$, the SN~Ia delay 
time must be {\it lower} than $2$~Gyr. If so, 
then the Lanzetta et al. (2002) SFH will be ruled out by the data
presented here.

\end{abstract}


\section{Introduction}

The evolution of the SN rate over cosmic
time is a key ingredient for understanding the chemical enrichment
history of the Universe and the formation and properties of galaxies 
and clusters. Since the progenitors of core-collapse SNe are massive, 
short-lived stars, 
their cosmic rate evolution will closely follow the universal
star formation history (SFH). On the other hand, 
the actual route that leads a white dwarf (WD) to explode 
as a SN~Ia is still an open question 
(see, e.g., Yungelson \& Livio 2000, and references therein).     

The mechanism that
leads to the explosions of SNe~Ia likely involves a significant delay between
the formation of the progenitor system and the explosion of the SN. The
rate of SNe~Ia at a given epoch is therefore
expected to depend on the star formation
rate up to several Gyr preceding that epoch. Thus, the 
cosmic evolution of SN rates can be used to constrain both the global SFH, 
and the characteristic delay time between
star formation and SN~Ia explosion (e.g., Madau, Della Valle \& Panagia 1998;
hereafter MDP). Since competing SN~Ia models predict different delay times, 
observational constraints on the delay can discriminate between 
some of the proposed scenarios for SN~Ia progenitor systems.

During the last few years, SNe~Ia have been used successfully as cosmological
distance indicators (e.g., Riess et al. 1998,2001; Perlmutter et al. 1999;
Tonry et al. 2003). Perhaps the main uncertainty that still
plagues SN-Ia-based distance measurements is the possibility of evolution in SN~Ia
properties. The use of SNe~Ia as distance estimators relies on the assumption 
that distant SNe~Ia at redshifts as high as $z=1.7$ are similar to local 
events. In order to study this question from a theoretical perspective, we
require knowledge on the nature of the progenitor systems of SNe~Ia. For any
given scenario, it is possible to estimate the evolution of high-$z$ systems
relative to local ones, e.g., as a result of metallicity or stellar-age
effects. Determination of SN~Ia delay times, with the resulting constraints
on progenitor models, are therefore desirable.

Motivated by these questions, a number of authors have recently studied the 
evolution of SN rates. Most studies have combined
models of the SFH with a recipe for the 
delay function of SNe~Ia, and have calculated the expected evolution
of SN rates, either per unit comoving volume, or per unit stellar luminosity
(e.g. J$\o$rgensen et al. 1997; Sadat et al. 1998; MDP; Ruiz-Lapuente \& Canal 1998; 
Yungelson \& Livio 1998; Kolatt \& Bartelmann 1998). 
As measurements of SN~Ia rates at high$~z$ have became available
(Pain et al. 1997; 2002) attempts have been made to compare the observed rates
with model predictions (e.g., MDP; Sadat et al. 1998; Kobayashi et al. 1998; 
Kobayashi, Tsujimoto, \& Nomoto 2000; Pain et al. 2002; Calura \& Matteucci 2003). 
However, no strong conclusions have been reached regarding the SFH or SN~Ia
progenitors. 

All available measurements
of SN rates beyond the local Universe (Pain et al. 1997; 2002; Hardin et 
al. 2000; Gal-Yam, Maoz, \& Sharon 2002; Tonry et al. 2003) are based on
SNe~Ia in the redshift range $z=0-1$. Due to the
limited number of observed SNe~Ia, these studies have distributed their 
samples into wide redshift bins. In
Gal-Yam et al. (2002) we divided our sample into two redshift bins, 
with $<z>=0.25$ and $<z>=0.9$. This binning caused no loss
of information in our work, as the number of bins was
comparable to the number of SNe ($2-3$). On the other hand,
Pain et al. (1997; 2002) and Tonry et al. (2003) used only one bin, and 
calculated the SN rate at an average redshift, $z\sim0.5$. In these cases, 
most of the redshift information is lost
in the binning process. 

This loss may be averted if, instead of deriving absolute rates from the
observations, and comparing them to predicted rates, we
begin with the predictions and fold them through the observational filters.
We can then compare the prediction for a particular experiment 
with the unbinned observations. 
Specifically, we can test whether or not the unbinned distribution of SN 
redshifts in a particular survey is consistent with some model.
Indeed, several authors have calculated such distributions
(e.g., Dahl\'en \& Fransson 1999, hereafter DF; Sullivan et al. 2000a). DF
noted that these distributions can be used to constrain 
progenitor models. However, they concentrated on future, deep
surveys with the {\it James Webb Space Telescope}, and did not
attempt to constrain model parameters using existing SN data.  

Pain et al. (1997, 2002) and Tonry et al. (2003) have calculated the
expected redshift distributions of SNe in their respective surveys, and have
compared them with the observed distributions as part of their derivation of 
SN rates. However, the SN redshift distributions calculated by Pain et al. 
and Tonry et al. are based on particular assumptions about the evolution of
the SN~Ia rate between $z=0$ and $z \sim 1$, which is assumed to be either
constant (Tonry et al. 2003), or to vary as some power of the redshift 
(Pain et al. 2002). The comparison between the expected and observed SN~Ia 
redshift distributions is used to find the best fitting rate evolution, and
to derive the average SN~Ia rate at the mean redshift ($z \sim 0.5$ in both
cases). Again, the information contained in the redshift distribution of
SNe~Ia at $z=0 - 1$ is lost in the averaging process. Both 
groups note that their data seem to indicate a slowly varying SN~Ia rate 
in this redshift range. Tonry et al. conclude that,
given that the slope of the SN~Ia rate is shallower than that of
some determinations of the SFH in this redshift range,
the typical delay time between star formation and 
SN~Ia explosion must be $\sim 1$ Gyr. 

In view of the fact that some models (e.g., MDP) predict strong evolution in
the SN~Ia rate at $z=0 - 1$, in contrast with the trends seen by 
Pain et al. (2002) and Tonry et al. (2003), we have undertaken a more
comprehensive investigation of the SN~Ia redshift distribution in this
range. As we will show below, existing data can already place
interesting limits on the SFH and on the SN~Ia characteristic delay time.

Throughout this paper, we assume a flat cosmology with $\Omega_m=0.3$
and $\Omega_{\Lambda}=0.7$. We designate with $h$ the Hubble parameter in 
units of 100 km s$^{-1}$ Mpc$^{-1}$. 

\section{Calculation of the Supernova Redshift Distribution}

We begin by calculating 
$N(z,{R_{lim}})$, the redshift distribution of SNe visible in a region
of sky at a given moment, to a limiting $R$-band magnitude $R_{lim}$.
Throughout this paper, we assume that all apparent magnitudes
are measured in the $R$-band. Our calculations can be applied
to any other band, as long as all apparent magnitudes are measured in the
same band. Then,
\begin{equation}
N(z,{R_{lim}}) = n(z) \times dV(z) \times T(z, R_{lim})~~~,
\label{eq1}
\end{equation}
\noindent where $n(z)$ is the SN rate density, defined as the
number of SNe per unit time per unit volume as a function 
of the redshift, and $dV(z)$ is the volume between $z$ and $z+dz$ 
in that region of sky. 
$T(z, R_{lim})$ is the effective visibility time of a SN at redshift $z$,
given the detection efficiency as a function of magnitude 
(this is often referred to as the ``control time''). $T$ depends on
$z$ because SNe at higher $z$ are fainter, and 
therefore can be detected for a shorter period above 
the limiting magnitude $R_{lim}$. 
The brightness of the SN in the chosen bandpass also depends on $z$
since different parts of the SN spectrum are redshifted into the 
observed band for SNe at different redshifts. 
 
The SN~Ia rate density, $n(z)$, is a convolution of the star-formation
history (SFH) with a ``delay'' or ``transfer function'', which is the SN~Ia 
rate vs time following a brief burst of star formation.
The delay function accounts for the time span between
star formation, through stellar evolution, the formation of WDs, 
and the final stage where a WD accretes material 
from (or spirals in and merges with) a binary companion (MDP; DF).
We follow here the delay function parameterization given by MDP. 
The overall time delay
includes the mass-dependent lifetime of the progenitor as a main-sequence
star, $\Delta t_{MS}$. Once the progenitor has gone off the main-sequence and 
has become a WD, it has a probability $\propto \exp(-{{\Delta t}\over \tau})$
to explode as a SN~Ia, where $\Delta t$ 
is the time since the star left the main sequence. DF have proposed a simpler model,
with a discrete delay time, $\tau$, which we further consider in section $\S~4.2$. 
Since we will be interested solely in 
the redshift distribution of SNe, we will ignore the normalization of $n(z)$. 
Integrating the explosion probabilities over the stellar initial mass
function, $dN/dm$, and the past SFH, $\Psi(t)$, the SN rate density 
as a function of cosmic time is 
\begin{eqnarray}
n(t) \propto \int^t_{0} \Psi(t')dt' \times \nonumber
\end{eqnarray}
\begin{equation}
\int_{m_{\rm min}(t-t')} 
^{m_{\rm max}} \exp(-{t-t'-\Delta t_{MS}\over \tau})~{{dN} \over {dm}}(m)~dm~~,
\end{equation}
\noindent where, $m_{min}$ and  
$m_{max}$ are the minimum and maximum initial masses that
will lead to the formation of a WD that explodes as a SN~Ia.  
Following MDP, we adopt 
\begin{eqnarray}
m_{\rm min}={\rm max}[3 M_{\odot}, 
({{t-t'}\over{10~{\rm Gyr}}})^{-0.4} M_{\odot}],~~ 
m_{\rm max}=8 M_{\odot},~~\nonumber
\end{eqnarray}
\begin{equation}
{\rm and}~~{{\Delta t_{MS}} \over {10~{\rm Gyr}}}=({{m} \over {M_{\odot}}})^{-2.5}.
\end{equation}
A Salpeter (1955) initial mass function (IMF), 
$dN/dm \sim m^{-2.35}$, is assumed. We convert $n(t)$ to $n(z)$ using
the transformation between cosmic time and redshift,
\begin{equation}
\Delta t = {1 \over H_{o}} \int^{z_2}_{z_1} {dz \over {(1+z)[(1+z)^3\Omega_{m} + \Omega_{\Lambda}]^{1/2}}}~~~.
\end{equation} 
  
The detailed form of
the SFH, $\Psi$, is a much debated issue in recent years. Initial
studies of the UV to IR luminosity density of star-forming galaxies (e.g., 
Lilly et al. 1996; Madau et al. 1996; Connolly et al. 1997, Madau, Pozzetti, \& 
Dickinson, 1998) suggested that the SFH increases sharply 
between redshifts of zero and $z\sim 1-2$, and then decreases
at higher $z$. However, this picture has been challenged.
For instance, Steidel et al. (1999) argue for an almost-constant 
SFH at $z\sim1 - 4$, while Lanzetta et. al. (2002) favor a
monotonic increase in SFH out to $z=10$. At lower redshifts, Cowie, Songaila, 
\& Barger (1999) and Sullivan et al. (2000a)
have found the SFH slope at $z \sim 0 - 1$ to be quite
shallow, $\Psi(z) \sim (1+z)$, but recent measurements by 
Hippelein et al. (2003) seem to confirm the sharp rise, 
$\Psi \sim(1+z)^4$, initially reported by Lilly et al. (1996). 

To parameterise this range of possibilities for $\Psi(z)$, we
represent the SFH by a broken power law, with\\
$\Psi~(z)~\sim~(1+z)^{\alpha}$ at high$~z$, and 
$\Psi (z) \sim (1+z)^{\beta}$ at low$~z$.%\footnotemark[1] 
%\footnotetext[1]{We note that in a recent paper by Baldry \& 
%Glazebrook (2003) they independently adopt an almost identical approach,
%including both the functional form and the ranges of the 
%$\alpha$ and $\beta$ indices.}
These functions are joined smoothly at $z=1.2$, using the prescription 
of Beuermann et al. (1999; see also Bersier et al. 2003), 
\begin{equation}
\Psi(z)={2^{1/s} \times \Psi(1.2) \over [({2.2 \over 1+z})^{\alpha \times s}
+({2.2 \over 1+z})^{\beta \times s}]^{1/s}}~~~,
\end{equation}

\noindent with $s=5$. 
The high-$z$ index $\alpha$ assumes values between $\alpha=-2$, 
which corresponds to the results of Madau et al. (1998), and $\alpha=2$, 
which describes the work by Lanzetta et al. (2002). Similarly, the low-$z$ index 
$\beta$ varies between $\beta=4$, as advocated by Hippelein et al. (2003), and 
$\beta=1$, tracing the results of Cowie et al. (1999)\footnotemark[1]. 
\footnotetext[1]{In the case of $\beta=1$, the Beuermann prescription
causes the high-$z$ portion of the SFH function to remain somewhat affected 
by the low-$z$ index $\beta$ even at redshifts much larger than $z=1.2$. 
In that case, we therefore use a simple broken power law.}
Fig. 1a shows two examples of the SFH parameterization. 
With these ingredients, each model for the SN rate density $n(z)$ 
is determined by three parameters - the typical delay time $\tau$ and the 
two indices describing the SFH, $\alpha$ and $\beta$. Fig. 1b shows two 
examples of SN~Ia delay functions, and Fig. 1c shows two examples of $n(z)$
for particular combinations of SFH and $\tau$. 

\begin{figure*}
\centerline{\epsfxsize=140mm\epsfbox{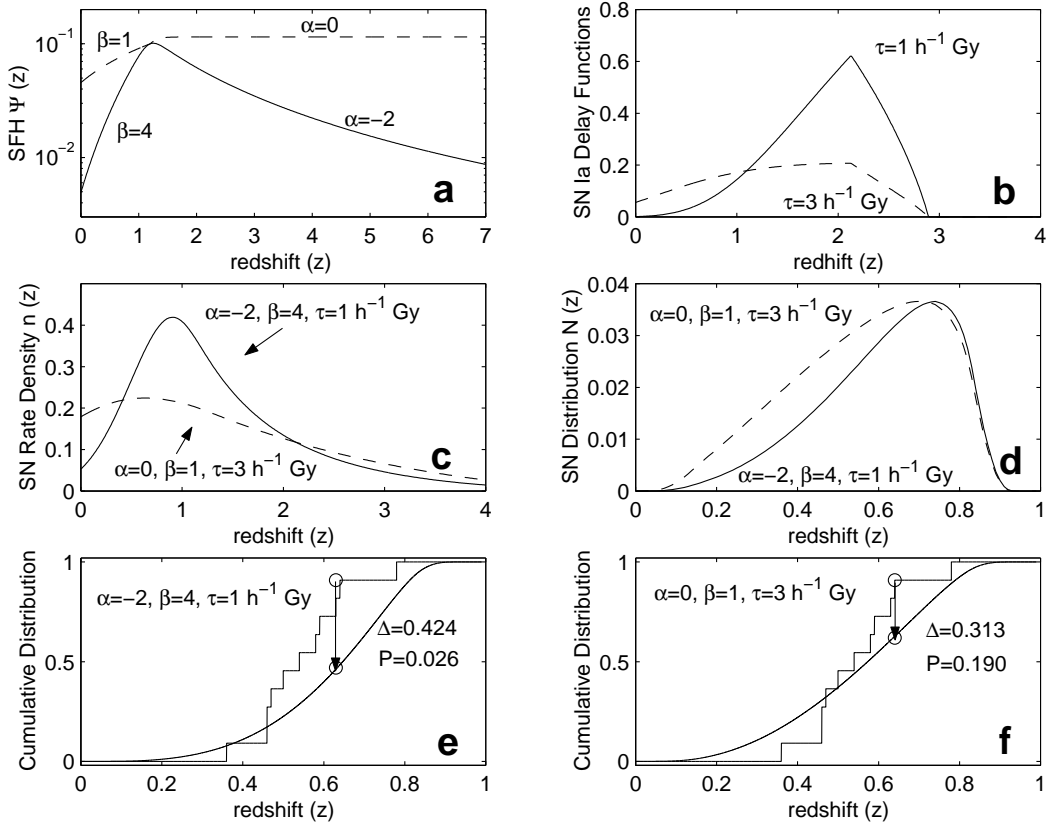}}
\caption{Illustration of the modeling and comparison to data. 
Panel a shows two examples of the SFH, $\Psi(z)$ - 
a ``Madau'' SFH, with a peak at $z=1.2$ (solid curve), 
and a shallower model (dashed curve) reflecting the proposed modifications 
by Cowie et al. (1999) and Steidel et al. (1999). 
Panel b shows two examples of 
the expected SN~Ia rate density following a brief
burst of star formation. These delay functions are 
calculated using the MDP prescription, with characteristic exponential 
delay times of $\tau=1 h^{-1}$ Gyr (solid) and $\tau=3 h^{-1}$ Gyr (dashed). 
For display purposes, an arbitrary redshift of $z=3$ has been chosen for
the burst of star formation.
SFH models are convolved with a delay function, and the
resulting SN rate densities $n(z)$ for a ``Madau'' SFH with $\tau=1 h^{-1}$ Gyr
(solid) and a ``Cowie-Steidel'' SFH with $\tau=3 h^{-1}$ Gyr (dashed) 
are shown in panel c. Panel d shows the predicted 
SN distributions, $N(z)$, for the models of panel c,
in a survey with the same observational parameters of the SCP search described 
in $\S~3$. KS tests show that the cumulative version of $N(z)$ from
a model combining a ``Madau'' SFH with a typical delay time of $\tau=1 h^{-1}$
Gyr (panel e) is ruled out by the data,  while a model with
``Cowie-Steidel'' SFH and $\tau=3 h^{-1}$ Gyr is consistent with the data
(panel f). Vertical axis units are arbitrary in panels a-d.}
\end{figure*} 

For our chosen cosmology, the volume element is given by,
\begin{equation}
dV(z) \propto D_{A}^2  \times c~ {dt \over dz} 
\propto {{[\int_0^z \chi dz']^2} \over {(1+z)^2}} \times
{\chi \over {(1+z)}}~~~,
\end{equation}
\noindent where $D_{A}$ is the angular diameter distance, and $\chi$ is given by
\begin{equation}
\chi = {1 \over {[\Omega_m (1+z')^3 + \Omega_{\Lambda}]^{1/2}}}~~~.
\end{equation} 

In the calculation of the effective visibility time $T(z,R_{lim})$ we follow 
the procedures described in Gal-Yam et al. (2002). 
Briefly, assuming initially that SNe~Ia are perfect standard candles with identical
light curves, the peak magnitude
of a SN at a given redshift is ${R_{peak}}(z)$, and 
the period it spends above some limiting magnitude ${R_{lim}}$
can be easily determined.  
The efficiency of a survey, 
$\eta$, is generally a function of the apparent magnitude 
$R$ (with $R_{lim}$ defined by $\eta(R_{lim})=0$), 
and is assumed to be independent of the SN 
redshift, i.e., the detection probability is the same for SNe with the same
magnitudes but different redshifts. 
Weighting the time the SN spends above
the detection limit by the relevant efficiency values, gives 
\begin{equation}
T(z) = \int_{0}^{\infty} \eta[R(t,z)] dt~~~,
\end{equation}
\noindent where $R(t,z)$ is the light curve of a SN~Ia at redshift $z$, and
$\eta=0$ when $R(t) \ge R_{lim}$. 
Note that cosmological time dilation will slow down the evolution
of the light curves (e.g., Leibundgut et al. 1996, Goldhaber et al. 1997, 
Riess et al. 1997), and hence lengthen the
visibility times of SNe~Ia by $(1+z)$, but this is canceled out
in the calculation of $N(z)$
by the $(1+z)$ reduction of the observed SN rate at redshift $z$.

The SN survey data whose redshift distribution we will model were obtained
in the $R$ band, and range in redshift from $z=0.361$ to $z=0.778$. We
require information on $R(t)$ and $R_{peak}(z)$ in order
to calculate $T(z)$. To determine $R_{peak}(z)$,
we fit a 3rd-order polynomial to the observed $R$-band 
peak magnitudes of high-$z$ SNe reported by Perlmutter et al. (1998, 1999).
Note that no K-corrections are needed,
as the input we require are the peak $R$ magnitudes at all
redshifts, and these do not have to be translated into rest frame
$B$ magnitudes, as done by the SCP for Hubble-diagram purposes. In terms
of the light curves, $R(t)$, the observer-frame $R$ band is a fair 
match for rest-frame $B$ band in most of the redshift range we model. 
However, at the highest and lowest 
redshifts, rest-frame $U$ and $V$ wavelengths, respectively, dominate 
the observer-frame $R$ band.
Since the light curves of SNe~Ia have slightly different forms 
in each band, we have constructed a set of $z$-dependent observer-frame
light curves, by linearly interpolating between the appropriate rest-frame 
light curves. The template light curves were adapted 
from $U$, $B$ and $V$ light curves, kindly supplied by
B. Leibundgut, and supplemented with data 
points taken from Riess et al. (1999), which also give
slightly better coverage of the 
pre-maximum period. Early $U$-band observations of SNe Ia are scarce, so we
had to extrapolate the rest-frame $U$-band light curve. 
However, this extrapolated curve is
only relevant for high-$z$ SNe, which are invariably 
discovered close to the limiting magnitude. Therefore, the influence 
of the extrapolated part of the light curve, well below the peak magnitude, 
is negligible. The $R_{peak}(z)$ curve and observer frame $R$-band light curves 
represent the properties of an average SN~Ia. 

In reality, the peak magnitudes and light curve shapes are not uniform. 
``Peculiar'' SNe~Ia are quite common in low-$z$ SN samples, but they are
apparently absent from high-$z$ samples like the ones we consider 
(Li et al. 2001). One may worry that many SNe~Ia of the ``underluminous'',
1991bg-like, variety are missed at high~$z$ due to their lower luminosity,
and we return to consider this possibility in section $\S~4.2$. Even
spectroscopically ``normal'' SNe~Ia are not perfect standard candles, 
and exhibit an intrinsic scatter of $0.2-0.3$ mag in peak luminosity (see 
Branch 1998, for a review). The peak magnitude of a SN~Ia is correlated with 
the light-curve shape of the object, with brighter events having a slower 
rise to maximum followed by a slower decline (e.g., Phillips 1993; 
Phillips et al. 1999; Riess, Press, \& Kirshner 1996; Riess et al. 1998; 
Perlmutter et al. 1995, 1999). A correction factor can be applied to the peak
magnitude to calibrate SNe~Ia as standard candles.
The intrinsic scatter in peak magnitudes 
and the differences in light curve shapes may cause a scatter in the visibility 
times of SNe, since brighter and broader SNe will have longer visibility times. 

Perlmutter et al. (1999) show the distribution of correction 
factors (their Fig. 4). In order to account for this distribution, we have 
calculated the effective visibility times for SNe with various peak magnitudes
and light-curve shapes. We then average these times using
the distribution of peak magnitudes
given by Perlmutter et al. (1999) as a weighting function.
We thus obtain a function $T(z,R_{lim})$ that fits a population of SNe~Ia with
properties similar to those observed by the SCP. 

Having determined $T(z,R_{lim})$ for a particular survey, we can
calculate the redshift distribution of SNe~Ia, $N(z)$, as
a function of the model parameters. Figure 1d 
illustrates $N(z)$ for the two examples of $n(z)$ in Figure 1c. 
We then convert these redshift distributions to cumulative form, and use the
Kolmogorov-Smirnov (KS) statistic to test if a data set 
is consistent with the predictions of a
specific model. Figure 1 (panels e-f) shows an example of this procedure.    
Bloom (2003) presents a similar treatment of the
redshift distribution of gamma-ray bursts.
  
\section{Application to an observed sample of high-$z$ Supernovae~Ia}

We now choose a sample of SNe~Ia whose redshift distribution can be
compared with predictions.    
Over the last few years, hundreds of high-$z$ SNe~Ia have been discovered, 
mainly by the SCP (Perlmutter et al. 1995) 
and the High-$z$ Supernova Search Team (Schmidt et al. 1998). These
SNe were discovered during numerous observing runs, using different
telescopes and detectors, through various pass-bands and under non-uniform 
conditions. Given a complete description of the
observations, a combined analysis of all available data sets should be feasible.
At present, we limit ourselves to an analysis of a single data set 
obtained with a uniform observational setup, and for which most of the observational 
details are available, in particular estimates of the survey efficiency $\eta$. 

The SCP have published the results of their efficiency studies for several SN 
samples (Pain et al. 2002). 
Here, we study the deepest and largest of these data sets. 
This SN sample is denoted as set ``D'' in the analysis of Pain et al.
(2002). Following is a brief outline of the search procedure used to discover
these SNe. Deep (3600 s, total) $R$-band exposures were obtained with the BTC
camera mounted on the 4m Blanco telescope at CTIO, on December 28-29, 1997.
These were compared with similar images obtained 36-37 days earlier. Seventeen 
candidate SNe were discovered, with discovery magnitudes $21.6 < R < 24.5$ and
redshifts $0.36 < z < 0.86$. Twelve SNe were spectroscopically confirmed as SNe~Ia. 
However, only eleven of these were retained in the ``statistical sample'' used by 
Pain et al. (2002) to calculate SN rates, and these constitute 
the sample we analyze, presented in Table 1.

\begin{table}
\caption {The SCP SN sample} \label {SN table}
\vspace{0.2cm}
\begin{centering}
\begin{minipage}{70mm}
\begin{tabular}{cc}
\hline
SN & Redshift\\ 
\hline
1997el & 0.636 \\
1997em & 0.460 \\
1997ep & 0.462 \\
1997eq & 0.538 \\
1997er & 0.466 \\
1997et & 0.633 \\
1997eu & 0.592 \\
1997ex & 0.361 \\
1997ey & 0.575 \\
1997ez & 0.778 \\
1997fa & 0.498 \\
\hline
\end{tabular}
\end{minipage}
\end{centering}
\end{table}

The SN sample was discovered 
in 11 different fields (Pain et al. 2002) observed under non-uniform 
atmospheric conditions, leading to survey efficiency curves $\eta(R)$ which
vary from field to field. We have averaged the efficiency curves measured 
for each field (kindly provided by R. Pain), 
to get the mean efficiency of the
survey, $\eta (R)$ (Figure 2). 
Since the efficiency enters our calculations in 
a linear fashion, calculating the SN redshift distribution using the 
mean curve is equivalent to performing
the calculation for each field individually (with the appropriate 
efficiency curve) and summing the results. We neglect a possible 
redshift dependence of the efficiency due to the variation in
the contrast between SN light and underlying host galaxy emission
within a constant aperture. This
contrast is reduced with redshift $\propto D_L^{-2}~(1+z)^4$, where
$D_L$ is the luminosity distance. For our chosen cosmology,
this gives a factor of $\sim3$ between $z=0.3$ and $z=1$. Galaxy 
and SN K-corrections will
offset some of this effect, and, from Fig. 2b of Pain et al. (2002)
we can estimate that the remaining correction to the SN detection efficiency
will be of a few per cent at most.  

\begin{figure}
\centerline{\epsfxsize=85mm\epsfbox{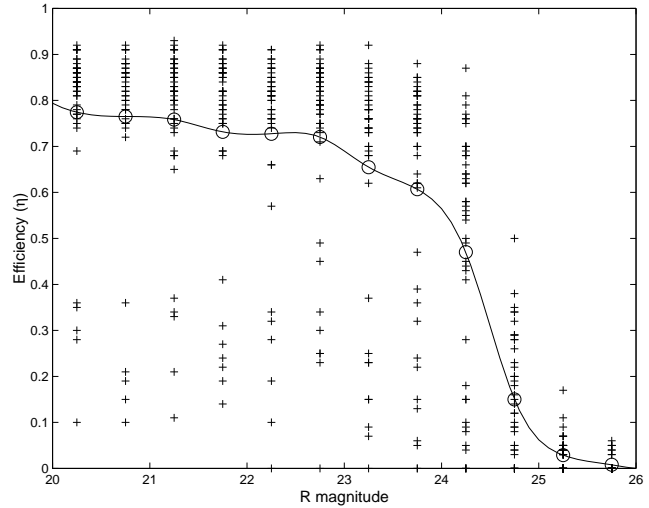}}
\caption{The mean efficiency of the SCP SN search described
in $\S~3$ as a function of $R$-band magnitude. Measurements
of the detection efficiency in individual fields, kindly supplied
by R. Pain, are given
by $+$ signs, while the open circles show the mean values.
The solid curve is a cubic spline fit to the mean points, and is used
in our calculation. Note the large
vertical dispersion resulting from non-uniform atmospheric 
conditions and from the variable fraction of each field that was
not used due to the presence of bright stars or other defects.}
\end{figure} 

The search strategy employed by the observers in a certain survey can also
influence the predicted redshift distribution of SNe~Ia. 
In particular, the SCP survey whose data we use here requires that a candidate
SN be brighter in a search image than in a reference image taken 
a few weeks earlier. 
This criterion was designed to discover only SNe near peak 
magnitude, and is driven by the SCP's 
scientific goals (Perlmutter 1999, and references therein). 
The search criterion has implications for the calculation of the effective
visibility times $T(z)$. For example, low-$z$ SNe, that are
much brighter than the survey's limiting magnitude $R_{lim}$, could,
in principle, contribute significantly to $T$ (Eq. 8).  
However, during most of the time these objects 
spend above $R_{lim}$, they are declining in magnitude, and therefore would
have been ignored by the SCP. We therefore limit the calculation of
$T(z)$ to the period during which a SN is brighter in the second epoch image
than it was 36 days before, when the reference images were obtained.
This is achieved by multiplying the integrand in Eq. 8 by 

\begin{equation}
S[R(t)]= \{^{1~~~R(t) < R(t-36 {\rm d})}_{0~~~R(t) \ge R(t-36 {\rm d})} 
\end{equation}

\section{Results}

\subsection{Star formation history and the delay time of SNe~Ia}

We have calculated the expected redshift distribution of SNe~Ia 
for the above survey, for a range of 
characteristic delay times $0.3 \le \tau \le 3~h^{-1}$ Gyr, and SFH 
indices $-2 \le \alpha \le 2$ and $1 \le \beta \le 4$.   
Examination of the full range of models calculated shows that the
data cannot constrain the model parameters individually, i.e., the
characteristic
delay time is degenerate with the SFH functional form. However, if we
assume some prior, either on the time delay $\tau$ or on the SFH, 
the data can be used to constrain the other parameters. For instance,
Figure 3 shows the range of allowed values for $\tau$ assuming
some of the SFH functions appearing in the recent literature. 
We can see that if the Madau et al. (1998) SFH is assumed,
the delay time $\tau$ is constrained to be $\tau \ge 1.7 (\tau \ge 0.7) 
h^{-1}$ Gyr at the $95\%(99\%)$ confidence level (CL). Rising SFHs, 
similar to those advocated by Lanzetta et al. (2002), are ruled out at 
the $95\%$ CL, unless $\tau > 3 h^{-1} $ Gyr.
A gentler evolution of the SFH, e.g, as proposed by
Cowie et al. (1999) and Steidel et al. (1999), places no constraints
on $\tau$. Inspecting the actual distributions predicted by the various
models (e.g., Fig. 1e and 1f), we find that invariably, 
the models that are ruled out by the data
under-predict the number of SNe~Ia at intermediate redshifts ($z \sim 0.5$)
relative to the number of SNe~Ia at higher redshifts ($z \sim 0.8$). 

\begin{figure}
\centerline{\epsfxsize=85mm\epsfbox{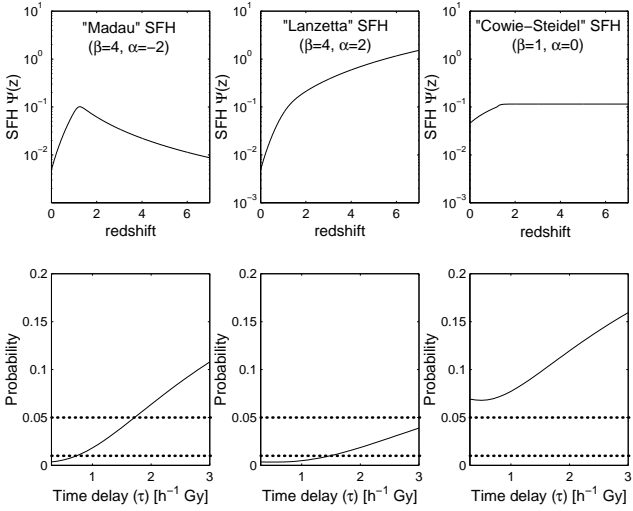}}
\caption{Probability of SN~Ia time delay values, given the data,
for particular SFH models. 
Assuming the SFH models shown in the upper panels, we can constrain
the allowed values of $\tau$ by the probability derived from the
KS test (lower panels). Points below the upper and lower dotted lines 
are ruled out at $95\%$ and $99\%$ confidence, respectively.}
\end{figure} 

\begin{figure}
\centerline{\epsfxsize=85mm\epsfbox{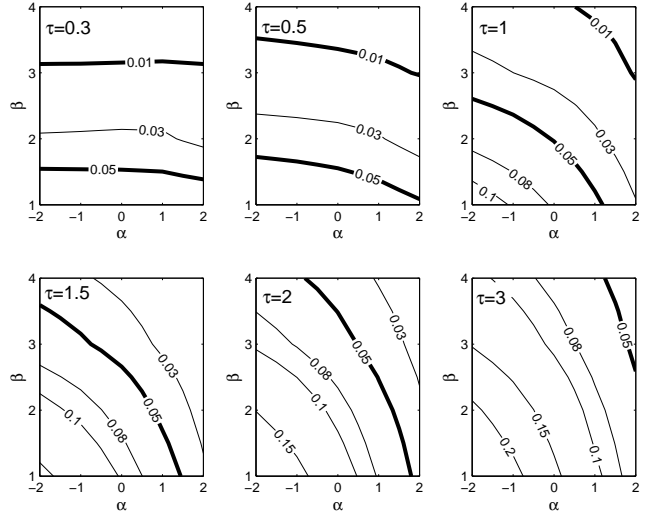}}
\caption{Constraints on SFH models for a given SN-Ia delay time.
Each panel shows contours of equal probability, derived from the
KS test for various combinations of the SFH indices $\alpha$ and $\beta$,
for the indicated SN~Ia delay time, $\tau$, in units of $h^{-1}$ Gyr.}
\end{figure} 

Figure 4 shows the constraints set on the SFH for given values of 
the delay time $\tau$. The first two panels show that, for short delay
times ($\tau<0.5 h^{-1}$ Gyr), the data constrain only the low-$z$ index
$\beta$. For such short delays, the SFH at high$~z$ does not
strongly affect (and is therefore not constrained by)
the SN redshift distribution we measure at $z<1$. 
If we assume such short time delays, the SN data require a gentle
decline in SFH from $z=1$ to $z=0$ ($\beta = 1$), favoring the 
results of Cowie et al. (1999) and Sullivan et al. (2000a) 
over those of Lilly et al. (1996) 
or the recent findings of Hippelein et al. (2003). 

As we consider longer time delays ($1 \le \tau \le 2 h^{-1}$ Gyr), the
data begin to constrain also the high-$z$ evolution of the SFH, 
parameterised by the index $\alpha$. Generally speaking, SFHs that decline
above $z=1.2$ (e.g., Madau et al. 1998) are favored over the strongly
rising functions of Lanzetta et al. (2002). For instance, a model where the 
SFH is rising at all $z$ as $\Psi \sim(1+z)^2$ (i.e., $\alpha=\beta=2$)
is ruled out at the 95\% CL for time delays $\tau \le 2 h^{-1}$ Gyr. 
The constraints are weakest when a long delay time, $\tau = 3 h^{-1}$ Gyr,
is considered, in which case only SFH models
combining a sharp increase at low redshift ($\beta > 2$) that continues
to rise steeply at high redshifts ($\alpha = 2$) are ruled out at the
95\% CL, and are thus disfavored by the data regardless of $\tau$. 
In conclusion, even with a small sample of 11 SNe~Ia, it appears
that the SN~Ia redshift distribution can be used to set interesting
limits on the SFH and the characteristic delay time of SNe~Ia.  

\subsection{Caveats}

We now consider possible caveats that might relax the constraints we have
found. These may be due to some weakness in the model we have employed, or to 
effects related to the data set we have used. 
We discuss both possibilities below.

Our model includes two main parameterised components -- 
the SFH and the SN~Ia delay function. Our treatment
of the SFH is fairly robust, since our parameterization 
covers the entire range of SFH functions reported in the literature. 
As for the SN~Ia delay function, 
if SNe~Ia are descended from a single or dominant type of progenitor, 
it will be reasonable to assume that there is a characteristic
time delay between the formation of these systems and the SN explosion. 
Our parameterised models
qualitatively reproduce the main features of most of the possible 
delay functions calculated for specific progenitor models by 
Ruiz-Lapuente \& Canal (1998) and Yungelson \& Livio (2000).  
A fastly rising and  then monotonically declining delay function
is also consistent with the observation that the rate of SNe~Ia 
appears to be higher among young stellar populations than among
older ones (e.g., Della Valle \& Livio 1994; Mannucci et al. 2003), 
However, the functional dependence could be quite different.

To further test the sensitivity of our results to the exact form of the
delay function, we have redone our calculations using the alternative 
parameterization
introduced by DF, who replace the exponential distribution of
explosion times used by MDP by a discrete value of the delay $\tau$.
The progenitor mass then uniquely determines the delay time
between the system formation and the SN~Ia explosion. The results of the 
calculations are shown in Figure 5. Comparing with Figure 4, we can see that
all the main findings we obtained using the MDP delay functions are reproduced,
and are in fact strengthened. As expected, because the delay time is
no longer given by an extended function, the SN redshift
distribution between $z \sim 0 - 1$ is influenced only by the low-$z$
SFH index $\beta$ for short delay times ($\tau<1 h^{-1}$ Gyr) and only
by the high-$z$ SFH index $\alpha$ for long delay times ($\tau \ge 2 h^{-1}$
Gyr). As before, for short delay times ($\tau \le 0.5 h^{-1}$ Gyr) the data
require a gently rising SFH ($\beta=1$). 
For longer delay times ($\tau \ge 1 h^{-1}$ Gyr)
we can rule out (at the $95\%$ CL) SFH models with a sharp rise at high~$z$ 
($\alpha = 2$) while the models involving also a sharp rise also at 
low~$z$ ($\alpha = 2$, $\beta > 2$) are ruled out at the $99\%$ CL regardless
of the time delay.  

\begin{figure}
\centerline{\epsfxsize=85mm\epsfbox{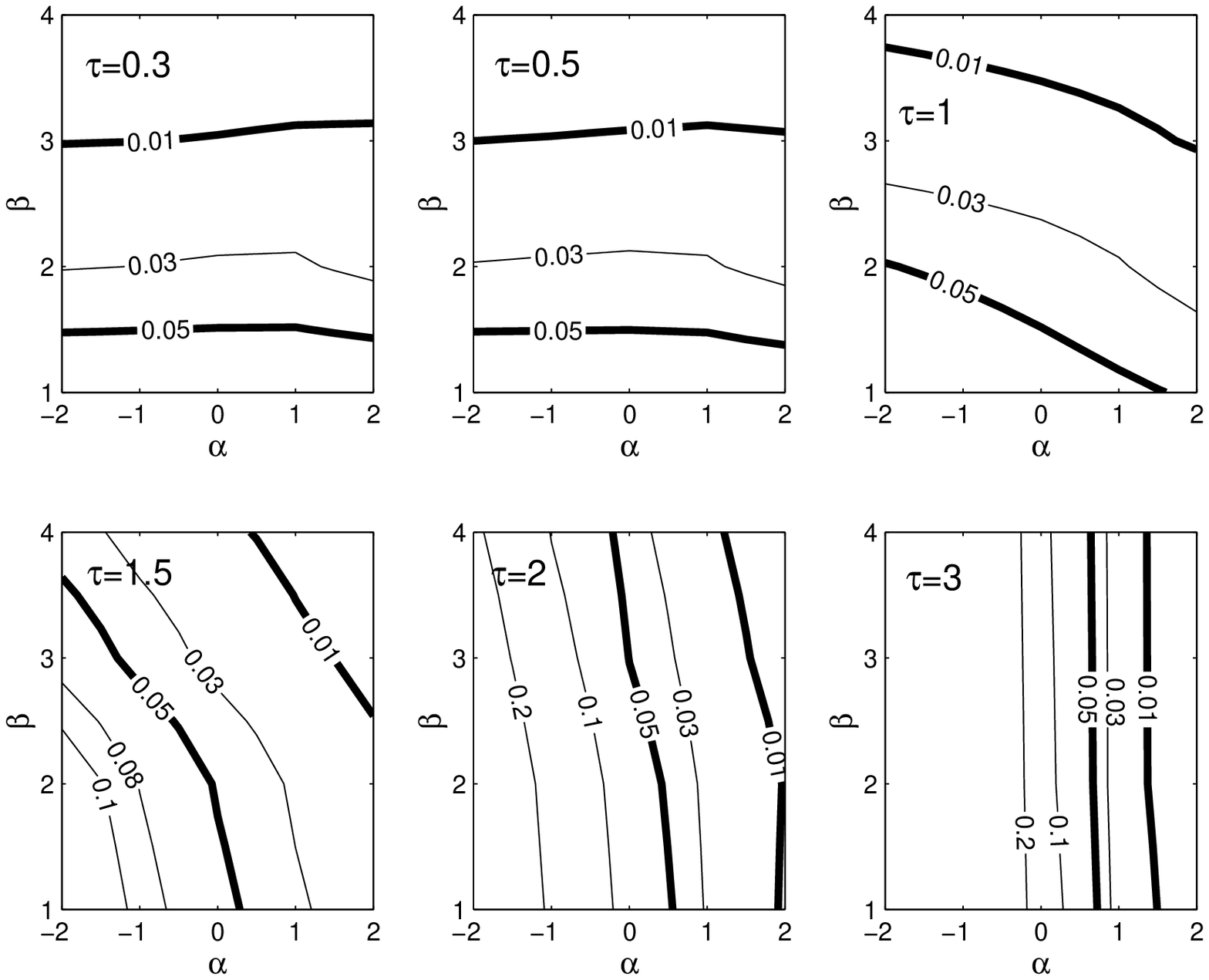}}
\caption{As in Fig. 4, but using the DF time delay parameterization 
($\S~4.2$).}
\end{figure} 

Several studies (e.g., Hamuy et al. 1995; Hamuy et al. 1996; Howell 2001) 
have suggested that some ``peculiar'' SNe~Ia (overluminous, 
SN 1991T-like, and underluminous, SN 1991bg-like) may result from distinct
subgroups of progenitors. It could therefore be argued that we should
consider also models using two or more characteristic delay times. 
However, it has been shown that high-$z$ SNe~Ia, which dominate the sample
we have considered, are at least as uniform, both spectroscopically 
and photometrically, as local ``normal'' SNe~Ia 
(Riess et al. 1998; Perlmutter 1999, and references within). 
Indeed, Li et al. (2001) report that both overluminous and underluminous SNe have 
not been detected at high$~z$ so far. Overluminous SNe would be over-represented in
high~$z$ samples. Their apparent absence probably rules our the possibility
that such events are common at high~$z$. 

To check what effect an undetected population of underluminous events
would have on our results, we
have constructed a set of SN light curves that have peak magnitude and light
curve shape similar to those of the prototypical underluminous 
SN~Ia, SN 1991bg. We then computed the effective visibility time using a mix
of SNe~Ia that includes $30\%$ 1991bg-like events, similar to the 
fraction found by Li et al. (2001) among local SNe~Ia. 
Due to their relative faintness, the redshift distribution of underluminous 
SNe~Ia peaks around
$z \sim 0.4$ (compared to $z\sim0.8$ for normal events, Fig 1d). We find that
this typically leads to an increase of $\sim 15\%$ to 
the total SN~Ia distribution at this redshift. We noted above that
the models calculated in section $\S~4.1$, and which fail to fit the data,
underpredict the number of SNe~Ia at 
intermediate redshift ($z \sim 0.4$) relative to those at higher redshifts.
Thus, the addition of SNe~Ia at intermediate redshifts due to an underluminous
SN population slightly weakens some the results reported above. For instance,
assuming a Madau et al. (1998) SFH now requires $\tau > 1.4 (0.6) h^{-1}$ Gyr at the
$95\%(99\%)$ CL, while a Lanzetta et al. (2002) SFH can be accomodated if 
$\tau > 2.5 (1.2) h^{-1}$ Gyr at the $95\%(99\%)$ CL. Other limits are not
relaxed -- short ($\tau < 0.5 h^{-1}$ Gyr) delay times require slowly
varying SFH at $z<1.2$ ($\beta = 1$), and steep, monotonically rising SFH
($\alpha=2, \beta \ge 3$) are ruled out at $95\%$ confidence regardless of $\tau$.       
Thus, an undetected population of underluminous high-$z$ SNe,
if it exists, does not significantly alter our results.   

With regard to the data set we have analyzed, one may speculate that perhaps 
the observers excluded some SNe that should have been included
in the sample, and may have thus affected the observed redshift distribution. However,
in their analysis, Pain et al. (2002)
investigated all variable sources that could have been SNe~Ia and which 
satisfied their search criteria. They determined whether or not
each candidate was a SN~Ia, based on spectroscopy and light curve shapes, 
leaving no unidentified 
sources. We have verified that the efficiency curves we have used
are consistent with the selection criteria used by Pain et al. (2002)
in compiling the SN sample shown in Table 1.

\subsection{Future Data Sets}

Finally, we consider potential applications of this analysis 
to future, larger, SN samples. We have created two simulated data sets.
The first simulated data set consists of  
100 SNe drawn from the expected distribution calculated for a survey with
limiting magnitude $R=24$, a ``Madau'' SFH ($\alpha=-2$, $\beta=4$), and
a typical time delay of $\tau=1 h^{-1}$ Gyr. A second sample of 1000 SNe from
a deeper survey to $R_{lim}=25$ mag is produced using
the same model parameters. We then treated each 
sample as real data and found constraints that could be drawn
on the $\alpha - \beta - \tau$ parameter space. Samples similar to
these synthetic sets are expected from ongoing or forthcoming
SN search programs such as ESSENCE (Smith et al. 2002; Garnavich et al.
2002), SNLS (Pain et al. 2002), SNAP (Perlmutter et al. 2002) and
the LSST (Strauss et al. 2002). 

Figure 6 shows results of the first simulation. A data set of 100 SNe
reaching $R=24$ mag should be a powerful tool to test the SFH and SN~Ia
delay times. While the input model ($\alpha=-2, \beta=4, \tau=1$) is
consistent with the data (P=0.49), significant portions of the
parameter space are ruled out. However, the degeneracy between the
SFH indices $\alpha$ and $\beta$, and the SN~Ia delay time $\tau$ persists,
and the data cannot constrain each parameter individually.        

\begin{figure}
\centerline{\epsfxsize=85mm\epsfbox{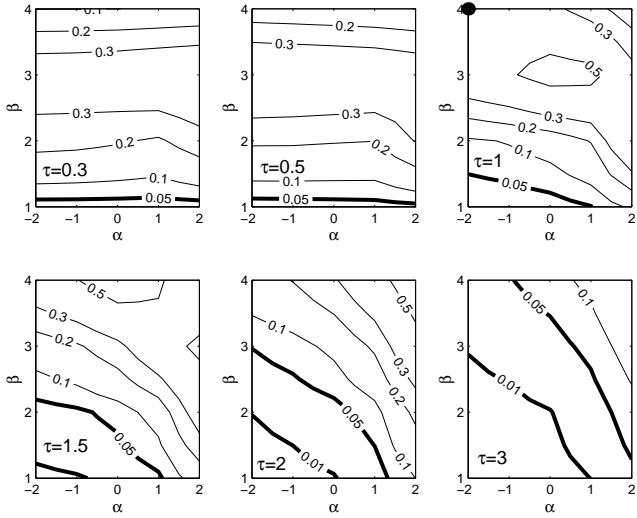}}
\caption{As in Fig. 4, but for a simulated data sample of 100 SNe,
drawn from the distribution calculated for $\alpha=-2$, $\beta=4$
and $\tau=1 h^{-1}$ Gyr and assuming a survey with limiting magnitude $R=24$.
Note that the input model (marked
with a filled circle) is recovered, and that significant areas of the
parameter space are ruled out. However, individual limits either on
the SFH indices ($\alpha$ and $\beta$) or the delay time
$\tau$ cannot be drawn.}
\end{figure} 

\begin{figure}
\centerline{\epsfxsize=85mm\epsfbox{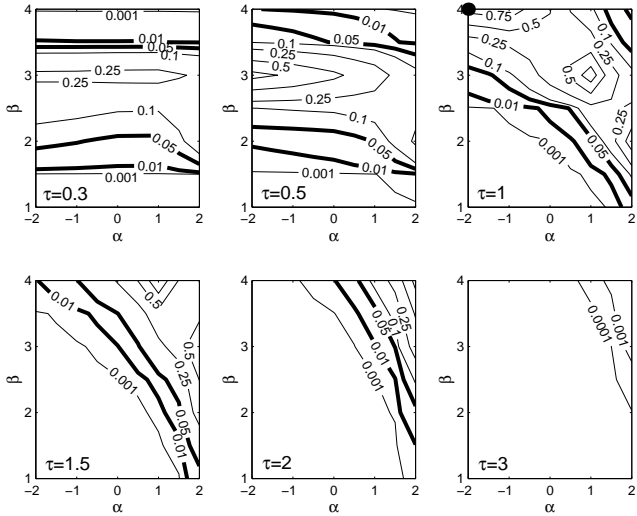}}
\caption{Similar to Fig. 6, but using a simulated sample of 1000 SNe
to a limiting magnitude $R=25$. Such a large data set 
could constrain both the SFH ($\beta > 1$ at the $95\%$ CL) and
the SN~Ia delay time ($\tau < 3 h^{-1}$ Gyr) simultaneously.}
\end{figure} 

Figure 7 shows the constraints that can be drawn from the larger, deeper
sample. The ``correct'' model ($\alpha=-2, \beta=4, \tau=1$)
is recovered with a probability value of $P=0.79$. Note that the degeneracy
between the form of the SFH and the SN~Ia delay time is now broken. Shallow
SFHs at low$~z$ $(\beta=1)$ and long delay times ($\tau=3$) are strongly 
ruled out by the data. Shorter delay times set strong constraints
on the allowed values of the SFH indices $\alpha$ and $\beta$.     

\section{Discussion and Conclusions}

We have calculated the expected redshift distribution of SNe~Ia in a magnitude
limited survey for a wide set of model parameters, and compared our
calculations with observations of high-$z$ SNe by the SCP. We have shown
that this approach is significantly more powerful than comparing 
expected and observed SN rates, because of the redshift binning involved 
in deriving SN~Ia rates. We confirm and quantify the findings of Pain
et al. (2002) and Tonry et al. (2003) which suggest that the redshift
distribution of SNe~Ia at $z \sim 0 - 1$ disfavor a strong variation
of the SN~Ia rate during this period. Thus, if the SFH between
$z=0$ and $1$ is steeper than $\Psi \propto (1+z)$, the delay time of
SNe~Ia must be long ($\tau > 0.5 h^{-1}$ Gyr). We show that, even for
longer delay times ($\tau \ge 1 h^{-1}$ Gyr), the data set interesting
constraints on the SFH. Generally, strongly rising SFHs, similar to
those advocated by Lanzetta et al. (2002), fit the data poorly, and 
SFH models which rise steeply at low redshifts ($\beta > 2$) and continue 
to rise at high redshifts ($\alpha=2$) are ruled out by the data
at the $95\%$ CL, unless $\tau > 3 h^{-1}$ Gyr. Turning the
argument around,    
given some determinations of the SFH (e.g., Madau et al. 1998), 
we can constrain the characteristic delay time of SNe~Ia to be 
$\tau > 1.7 h^{-1}$ Gyr ($\tau > 0.7 h^{-1}$ Gyr) at the $95\%(99\%)$ CL.    

As already noted, Yungelson \& Livio (2000) have studied the delay
functions of different 
SNe~Ia progenitor systems. In their double-degenerate (DD) model, 
two WDs merge, with the resulting WD reaching (or surpassing) the
Chandrasekhar mass. Single degenerate models include accretion 
of He-rich material from a non-degenerate companion, leading to 
helium ignition on the WD surface and an edge-lit detonation (He-ELD).
Alternatively, H-rich material from a main sequence companion is 
accreted and processed on the WD surface, and results either in an
edge-lit detonation due to accumulated helium (MS-ELD) or to
central carbon ignition as the WD reaches the Chandrasekhar mass
(MS-CH). 

The delay functions calculated for these models
are more complex than the parameterised MDP forms we have used, but generic
similarities exist. In particular, both DD and He-ELD curves 
can be characterised by a fast onset of SNe~Ia, some $3 \times 10^7$ years 
after star formation, followed by gradual increase and a decline 
that terminates in a sharp, exponential cutoff after $\sim 1$ Gyr 
for the He-ELD and $\sim 11$ Gyr for the DD models. This is qualitatively 
similar to the MDP exponential formulation we use. The MS-CH model
can be approximated by a DF-like model with a delay time of
$\sim 1$ Gyr, so can be constrained by our calculation using
this parameterization ($\S~4.2$, Figure 5). 
The MS-ELD is more complex, with a delayed 
onset of SNe~Ia, about $0.3$ Gyr after star formation, followed by an
approximately power-law decline lasting some $11$ Gyr.     

If we consider those SFH determinations requiring long time delays
in our models (e.g., Madau et al. 1998; Lanzetta et al. 2002), 
then the data disfavor the MS-CH and He-ELD progenitor 
models. We have shown that larger samples
of SNe~Ia, that will be obtained by future SN search
programs, some of which are already in progress, could be used to
obtain even stronger constraints on the SFH and SN~Ia models. 

In a companion paper (Maoz \& Gal-Yam 2003), we study the implications
of the measured SN~Ia rates in $z \sim 1$ galaxy clusters on 
the nature of SN~Ia progenitors and the source of iron in clusters.
We demonstrate there that, if SNe~Ia have produced most 
of the iron in galaxy clusters, and the stars in clusters formed 
at $z\sim2$, then the SN~Ia delay 
time must be {\it lower} than $2$~Gyr. Thus, if both conditions were met, 
then the Lanzetta et al. (2002) SFH would be ruled out by the data
presented here.

Finally, it is interesting to note here that most previous authors have considered
core-collapse SNe, whose rates closely track the SFH, as promising 
indicators for SFH measurement (but see DF
1999 for an alternative view). In principle, our methods could easily be
applied also to high-$z$ core-collapse SN samples, if and when they 
become available. However, recent work by
Mannucci et al. (2003) has shown that the majority of core-collapse
events in star-forming galaxies suffers from strong dust extinction. SFH
measurements using core-collapse SNe thus share the same problems 
that make UV-determined SFHs so widely debated -- the strong dependence
of the derived SFH on the little known amount and properties of dust at high redshifts. 
In contrast, most SNe~Ia probably occur in relatively dust-free environments.
Thus, measurements of the SFH using SNe~Ia, through methods similar
to the ones we have discussed here, may provide an attractive alternative
to SFHs based on UV and emission line fluxes.

\section*{Acknowledgments}

We thankfully acknowledge help, advice and fruitful discussions with
R. Pain, T. Dahl\'en, P. Madau, C. Porciani, S. Fabbro, S. Perlmutter, 
B. Leibundgut, A. Filippenko, B. Schmidt, E. Ofek, O. Gnat, D. Poznanski
and K. Sharon. This work was supported by the Israel Science 
Foundation --- the Jack Adler Foundation for Space Research, Grant 63/01-1.

\clearpage


\end{document}